\documentclass[11pt,a4paper]{article}

\usepackage{colortbl}
\usepackage{ascmac}
\usepackage{bm}
\usepackage[dvipdfmx]{graphicx}

\usepackage{amsmath}
\usepackage{amsfonts}

\usepackage{youngtab}

\def\tr{\mathop{\rm tr}\nolimits}
\def\Pexp{\mathop{\rm Pexp}\nolimits}
\def\diag{\mathop{\rm diag}\nolimits}
\def\vol{\mathop{\rm vol}\nolimits}

\begin{document}
\Yautoscale0
\newcommand{\YT}{\Yboxdim15pt\young}
\newcommand{\y}{\Yboxdim3pt\yng}
\newcommand{\Y}{\Yboxdim10pt\yng}
\Yvcentermath1

\begin{titlepage}
\title{
\vspace{-1.5cm}
\begin{flushright}
{\normalsize TIT/HEP-700\\ April 2024}
\end{flushright}
\vspace{1.5cm}
\LARGE{Brane expansions for \\ anti-symmetric line operator index}}
\author{
Yosuke {\scshape Imamura\footnote{E-mail: imamura@phys.titech.ac.jp}}
and
Masato {\scshape Inoue\footnote{E-mail: masatoinoue@phys.titech.ac.jp}}
\\
\\
{\itshape Department of Physics, Tokyo Institute of Technology}, \\ {\itshape Tokyo 152-8551, Japan}}

\date{}
\maketitle
\thispagestyle{empty}
\begin{abstract}
Based on the D5-brane realization of Wilson line operators in anti-symmetric representations,
we propose brane expansion formulas for $I_{N,k}$, the Schur index of ${\cal N}=4$ $U(N)$ SYM
decorated by line operators in the anti-symmetric representation of rank $k$.
For the large $N$ index $I_{\infty,k}$ we propose a double-sum expansion,
and for finite $N$ index $I_{N,k}$ we propose a quadruple-sum expansion.
Objects causing finite $k$ and finite $N$ corrections are disk D3-branes
ending on the D5-brane.
\end{abstract}

\end{titlepage}

\tableofcontents

\section{Introduction}
Wilson lines are the most basic
and important observables in gauge theories.
They serve as order parameters for detecting phase transitions.
They are also useful in gaining deeper insights into
phenomena such as dualities.
In this work, we study
Wilson line operators of ${\cal N}=4$ $U(N)$ SYM
within the framework of the AdS/CFT correspondence
\cite{Maldacena:1997re,Gubser:1998bc,Witten:1998qj}.

Wilson line operators are labeled by representations of the gauge group.
We consider $U(N)$ gauge theories,
and irreducible representations are specified by
partitions $\mu$ ($\ell(\mu)\leq N$), which are depicted as
Young diagrams.
The number of boxes $k=|\mu|$ in $\mu$ is referred to as the rank of the representation.

Different holographic realizations exist for Wilson lines.
Wilson line operators in the fundamental representation ($\mu=\{1\}$)
are realized by fundamental strings in the AdS space
\cite{Rey:1998ik,Maldacena:1998im}.
If the rank of the representation is of order $N$,
the coincident strings are more naturally described by D-branes.
A Wilson line in the symmetric representation with $\mu=\{k\}$
is related to a tubular D3-brane with a cross section $S^2\subset AdS_5$ \cite{Drukker:2005kx} (See also a discussion in Section \ref{disc.sec}).
For the anti-symmetric representation with $\mu=\{1^k\}$,
the corresponding object is a tubular D5-brane with a cross section $S^4\subset S^5$ \cite{Yamaguchi:2006tq}.
In both cases, the charge of $k$ fundamental strings is carried by the
electric flux on the D-brane.
If the rank is of order $N^2$,
the strings or D-branes deform
the background spacetime significantly, and
the line operators are naturally described by
bubbling geometries \cite{DHoker:2007mci}.

Among these realizations of line operators,
we specifically focus on the anti-symmetric line operators
realized by D5-branes.
We investigate them by computing the Schur index \cite{Gadde:2011uv},
which is a specialization of
the superconformal index \cite{Romelsberger:2005eg,Kinney:2005ej}.
See \cite{Bourdier:2015wda,Pan:2021mrw,Hatsuda:2022xdv} for analytic
formulas of the Schur index without line operator insertion.
These indices are regarded as
the partition function in $S^3\times S^1$,
and line operator indices
\cite{Dimofte:2011py,Gang:2012yr,Drukker:2015spa,Hatsuda:2023iwi,Guo:2023mkn,Hatsuda:2023imp,Hatsuda:2023iof}
are defined as the partition functions with insertion of BPS line operators.
We insert $1/2$ BPS Wilson line operators at the poles of $S^3$,
thereby breaking
the rotational symmetry $so(4)$ of $S^3$ to $so(3)$.
The R-symmetry $so(6)$ is also broken to $so(5)$.
Let $J_1$ be the Cartan generators of the $so(3)$,
and $R_x$ and $R_y$ be the
Cartan generators of the $so(5)$.
The Schur index of ${\cal N}=4$ SYM is defined using these generators by
\begin{align}
\tr[(-1)^Fq^{J_1}x^{R_x}y^{R_y}],\quad
q=xy.\label{indexdef}
\end{align}

We can regard the index with line operator insertion as a correlator of
the line operators in the spacetime $S^3\times S^1$, and we
use the notation $\langle\cdots\rangle^{U(N)}$.
We can calculate the index with line operators in a representation $\mu$ and
the cojugate of a representation $\nu$
inserted at the north and the south poles, respectively, by the localizatin formula
\begin{align}
\langle s_\mu(U)s_\nu(U^{-1})\rangle^{U(N)}
=\int_{U(N)}\hspace{-1.3em}dU\Pexp(i_{\rm vec}\chi_{\rm adj}^{U(N)}(U))
s_\mu(U)s_\nu(U^{-1}),
\label{localization}
\end{align}
where $i_{\rm vec}$ is the letter index of the ${\cal N}=4$ vector multiplet
\begin{align}
i_{\rm vec}=1-\frac{(1-x)(1-y)}{1-q},
\label{ivec}
\end{align}
and $\chi_{\rm adj}^{U(N)}(U)$ is the $U(N)$ adjoint character
\begin{align}
\chi_{\rm adj}^{U(N)}(U)=\tr(U)\tr(U^{-1}).
\end{align}
$\Pexp$ is the plethystic exponential,
and $\int_{U(N)}dU$ is the integral over the $U(N)$ gauge group manifold with the Haar measure.
$s_\mu(U)$ is the $U(N)$ character of an irreducible representation $\mu$.
For a diagonalized matrix $U=\diag(z_1,\ldots,z_N)$ $s_\mu(U)$ are symmetric polynomials of the $N$
gauge fugacities $z_a$ ($a=1,\ldots,N$), which are called Schur polynomials.
We will abuse this term to refer to $s_\mu(U)$, too.
On the left hand side in (\ref{localization})
we express the line operators by the corresponding Schur polynomials.

The line operator index we are interested in is
\begin{align}
I_{N,k}=\langle s_{\{1^k\}}(U)s_{\{1^k\}}(U^{-1})\rangle^{U(N)}\quad(0\leq k\leq N).
\label{inkdef}
\end{align}
It is well known that the large $N$ index without insertion, $I_{\infty,0}$,
is reproduced as the supergravity index on the AdS side \cite{Kinney:2005ej};
\begin{align}
I_{\infty,0}=I_{\rm sugra}=\Pexp\left(\frac{x}{1-x}+\frac{y}{1-y}-\frac{q}{1-q}\right).
\end{align}
For the insertion of fundamental line operators,
it was confirmed in \cite{Gang:2012yr} that
the index $I_{\infty,1}$ is reproduced by including the contribution from
the fluctuation modes on a fundamental string extended along $AdS_2$
analyzed in \cite{Faraggi:2011bb}.
\begin{align}
\frac{I_{\infty,1}}{I_{\rm sugra}}=I_{\rm F1}=\Pexp(x+y-q)=1+\frac{x}{1-x}+\frac{y}{1-y}.
\label{Ifund}
\end{align}

For the finite $N$ index $I_{N,0}$ without line operator insertion, the finite $N$ corrections are
reproduced by D3-branes wrapped around topologically trivial three-cycles in $S^5$,
called giant gravitons \cite{McGreevy:2000cw,Mikhailov:2000ya}.
The finite $N$ index can be given in the form of expansions
with respect to the number of giant gravitons, and such expansions
are called giant graviton expansions
\cite{Arai:2019xmp,Arai:2020qaj,Imamura:2021ytr,Gaiotto:2021xce}.
See also \cite{Murthy:2022ien,Lee:2022vig,Beccaria:2023zjw,Beccaria:2024szi}.

Such expansions were generalized to the line operator index
of the fundamental representation $I_{N,1}$
in \cite{Imamura:2024lkw}.
(See also \cite{Beccaria:2024oif}.)

The purpose of this paper is to propose similar holographic relations for $I_{N,k}$ with
finite $N$ and finite $k$ based on the
D5-brane realization of anti-symmetric line operators.

We will heavily rely on
numerical analysis in which we treat indices as Taylor series.
We use two different Taylor series
with respect to two different variables.
One is the $y$-series in the form
\begin{align}
\sum_{i=0}^\infty f_i(x)y^i,
\end{align}
and the other is the $t$-series in the form
\begin{align}
\sum_{i=0}^\infty f_i(u)t^i,
\end{align}
where $t$ and $u$ are defined by
the relations
\begin{align}
x=tu,\quad
y=tu^{-1}.
\end{align}
In the following we mainly use the variables $x$ and $y$.
When we consider $t$-expansion of a function, we implicitly
regard it as a function of $t$ and $u$, even if it is written
in terms of $x$ and $y$.

\section{Finite $k$ corrections}
\subsection{Large $N$ and large $k$ limit}
Let us first consider the large $N$ index $I_{\infty,k}$ given
by the formula
\cite{Hatsuda:2023iwi}
\begin{align}
\frac{I_{\infty,k}}{I_{\rm sugra}}
&=\sum_{\lambda\vdash k}\frac{1}{z_\lambda}[I_{\rm F1}]_\lambda.
\label{kpart}
\end{align}
(See Appendix \ref{characterexp.sec} for a derivation.)
$\sum_{\lambda\vdash k}$ is the
summation over partitions of an integer $k$,
and $[\cdots]_\lambda$ is defined by
\begin{align}
[f(x,y)]_\lambda=\prod_{i=1}^{\ell(\lambda)}f(x^{\lambda_i},y^{\lambda_i})
\end{align}
for a function $f(x,y)$ of fugacities $x$ and $y$.
$z_\lambda$ is defined for a partition
$\lambda=\{\lambda_1,\lambda_2,\ldots,\lambda_{\ell(\lambda)}\}
=\{\ldots,3^{r_3},2^{r_2},1^{r_1}\}$
by $z_\lambda=\prod_{m=1}^\infty m^{r_m}r_m!$.
The $t$-series expansions of $I_{\infty,k}/I_{\rm sugra}$ for small $k$ are
\begin{align}
I_{\infty,0}/I_{\rm sugra}&=1,\nonumber\\
I_{\infty,1}/I_{\rm sugra}&=1+2t+2t^2+2t^3+2t^4+2t^5+2t^6+2t^7+{\cal O}(t^8),\nonumber\\
I_{\infty,2}/I_{\rm sugra}&=1+2t+5t^2+6t^3+9t^4+10t^5+13t^6+14t^7+{\cal O}(t^8),\nonumber\\
I_{\infty,3}/I_{\rm sugra}&=1+2t+5t^2+10t^3+15t^4+22t^5+31t^6+40t^7+{\cal O}(t^8),\nonumber\\
I_{\infty,4}/I_{\rm sugra}&=1+2t+5t^2+10t^3+20t^4+30t^5+48t^6+68t^7+{\cal O}(t^8),\nonumber\\
I_{\infty,5}/I_{\rm sugra}&=1+2t+5t^2+10t^3+20t^4+36t^5+58t^6+90t^7+{\cal O}(t^8),\nonumber\\
I_{\infty,6}/I_{\rm sugra}&=1+2t+5t^2+10t^3+20t^4+36t^5+65t^6+102t^7+{\cal O}(t^8).
\end{align}
We showed unrefined index with $x,y\rightarrow t$ to save the space.
We can clearly see the convergence in the limit $k\rightarrow\infty$.

The analytic expression for the large $k$ limit is obtained as follows.
Notice that we can regard
the right hand side of (\ref{kpart}) as the partition function of
$k$ identical particles.
If we take the large $k$ limit with fixed energy, almost all particles are in the ground state (Bose-Einstein condensate),
and the number of excited particles is not constrained.
The ground state is given by the term ``$1$'' in (\ref{Ifund}),
and particles in the ground state do not contribute to the index.
Therefore, the large $k$ limit of the index becomes \cite{Gang:2012yr}
\begin{align}
\lim_{k\rightarrow\infty}\frac{I_{\infty,k}}{I_{\rm sugra}}
&=\Pexp(I_{\rm F1}-1)\nonumber\\
&=1+2t+5t^2+10t^3+20t^4+36t^5+65t^6+110t^7+{\cal O}(t^8).
\label{largek}
\end{align}
In the second line we again showed the unrefined index.

This result agrees with the index of the fluctuation modes on the D5-brane.
The insertion of large $k$ anti-symmetric line operators
is realized on the AdS side by introducing a D5-brane
extended along $AdS_2\times S^4$ \cite{Yamaguchi:2006tq}.
The radus of $S^4$ is given by $L_{\rm AdS}\sin\theta_{\rm D5}$ with the angle $\theta_{\rm D5}$
determined by the relation
\begin{align}
\frac{k}{N}
=\frac{1}{\pi}(\theta_{\rm D5}-\sin\theta_{\rm D5}\cos\theta_{\rm D5}).
\label{novern0}
\end{align}
If $k$ is of order $N$, the radius of $S^4$ is comparable to the
AdS radius $L_{\rm AdS}$, and we can treat the brane as
a semi-classical object.
Mode analysis of the fields on the D5-brane was performed
in \cite{Faraggi:2011bb,Faraggi:2011ge},
and the corresponding index
\begin{align}
I_{\rm D5}=\Pexp\left(\frac{x}{1-x}+\frac{y}{1-y}\right)
\label{id5}
\end{align}
agrees with 
(\ref{largek}).

\subsection{Full expansion}
In the large $k$ limit, the index is given by (\ref{largek}).
We are interested in the finite $k$ correction of ${\cal O}(t^{k+1})$ in the index $I_{\infty,k}$.
\begin{align}
\frac{I_{\infty,k}}{I_{\rm sugra}I_{\rm D5}}
=1+{\cal O}(t^{k+1}).
\end{align}
What are objects causing the finite $k$ correction on the AdS side?

\begin{figure}[htb]
\centering
\includegraphics{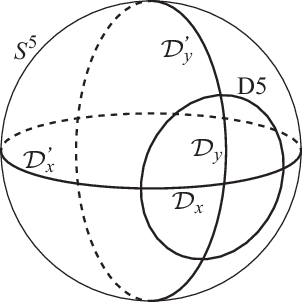}
\caption{The small circle is a cross-section ($S^4$) of the tubular D5-brane.
The $R_x$-fixed locus ${\cal C}_x$ shown as the horizontal circle is divided into
two parts ${\cal D}_x$ and ${\cal D}'_x$ by the D5-brane.
Similarly, the $R_y$-fixed locus ${\cal C}_y$ shown as the vertical circle
is devided into ${\cal D}_y$ and ${\cal D}'_y$.
Some important features are not shown correctly.
For example, ${\cal D}_x$ and ${\cal D}_y$ intersect not at a point but along a segment,
and ${\cal D}_x$ and ${\cal D}'_y$ are not separated as shown but in contact at two points.}\label{disks.eps}
\end{figure}

Let ${\cal C}_x$ and ${\cal C}_y$ be the
$R_x$-fixed locus and $R_y$-fixed locus in $S^5$, respectively.
They are homologically trivial $S^3$ in $S^5$.
In the giant graviton expansion of the Schur index without line operator insertion \cite{Arai:2020qaj},
D3-branes wrapped around ${\cal C}_x$ and ${\cal C}_y$ give finite $N$ corrections
to the index.
Let us define the latitude $\theta$ ($0\leq\theta\leq\pi$) in $S^5$
so that two poles $\theta=0$ and $\theta=\pi$ are located on the
intersection ${\cal C}_x\cap{\cal C}_y$.
The metric of $S^5$ is
\begin{align}
ds_{S^5}^2=L_{\rm AdS}^2(d\theta^2+\sin^2\theta ds_{S^4}^2).
\end{align}
A cross section of
the tubular D5-brane is $\theta=\theta_{\rm D5}$,
and the three-cycle ${\cal C}_x$ is divided by the D5-brane
into two parts $0\leq\theta\leq\theta_{\rm D5}$ and $\theta_{\rm D5}\leq\theta\leq\pi$,
which we denote by ${\cal D}_x$ and ${\cal D}'_x$.
Similarly, ${\cal C}_y$ is divided into two parts ${\cal D}_y$ and ${\cal D}'_y$.
(See Figure \ref{disks.eps}.)

A suggestive fact is that
the right hand side of (\ref{novern0}) is the ratio of the volumes
\begin{align}
\frac{k}{N}
=\frac{\vol({\cal D}_x)}{\vol({\cal C}_x)}
=\frac{\vol({\cal D}_y)}{\vol({\cal C}_y)}.
\label{volumeratio}
\end{align}
It is essentially important in the giant graviton expansion
that
the mass of a D3-brane wrapped around ${\cal C}_x$ and ${\cal C}_y$ is $N/L_{\rm AdS}$,
and the corresponding operator dimension is $N$.
(\ref{volumeratio}) means that the mass of 
a D3-brane on ${\cal D}_x$ and ${\cal D}_y$ is $k/L_{\rm AdS}$,
and this fact suggests that
such D3-branes
give the finite $k$ corrections to the index.

Based on analogy with giant graviton expansions for finite $N$ corrections to $I_{N,0}$,
we propose the expansion
\begin{align}
\frac{I_{\infty,k}}{I_{\rm sugra}I_{\rm D5}}=\sum_{m_x,m_y=0}^\infty x^{km_x}y^{km_y}{\cal F}_{m_x,m_y},
\label{finitek}
\end{align}
where $m_x$ and $m_y$ are the numbers of disk D3-branes on ${\cal D}_x$ and ${\cal D}_y$,
respectively.
In this section we consider the large $N$ limit with finite $k$,
and then $k'=N-k$ is infinite and D3-branes on ${\cal D}'_x$ and ${\cal D}'_y$ are
decoupled.
$k$-independent functions
${\cal F}_{m_x,m_y}$ are the indices of the theories realized
on the system consisting of $m_x$ D3-branes on ${\cal D}_x$ and
$m_y$ D3-branes on ${\cal D}_y$ (and the tubular D5-brane).

Because the branes causing the finite $k$ correction
are not giant gravitons but disk D3-branes,
we call (\ref{finitek}) (and similar expansions we will see below)
``a brane expansion'' rather than a giant graviton expansion.

We assume that the contribution from the D5-brane
is not affected by the introduction of disk D3-branes,
and in the analysis of D3-branes we treat the D5-brane
as a rigid object.
This is justified if the ratio of the tensions of the two branes
$T_{D3}/T_{D5}\sim l_s^2$ is much smaller than the typical scale $L_{\rm AdS}^2\sim l_s^2(Ng_s)^{1/2}$.
This is the case when the 't Hooft coupling $\lambda\sim Ng_s$ is large.
Even if it is not the case
the result should be correct because the index is independent of
continuous parameters.

Let us first consider the functions ${\cal F}_{m,0}$, which are the index of the theories realized on ${\cal D}_x$.
Such an index of theories living on a three-dimensional disk is known as the half-index \cite{Dimofte:2011py},
and for ${\cal N}=4$ $U(m)$ SYM it is given by
\begin{align}
\Pi_m
=\int_{U(m)}\hspace{-1.3em}dU
\Pexp(i_{\rm half}\chi_{\rm adj}^{U(m)}(U)),\quad
i_{\rm half}=\frac{x-q}{1-q}.
\label{pimdef}
\end{align}
Analytic form of the integral is\footnote{We did not analytically derive this analytic form,
but only checked the agreement of Taylor expansions up to a very high order.}
\begin{align}
\Pi_m
=\Pexp\left(
\frac{x^m-q}{1-q}+\frac{x-x^m}{1-x}
\right).
\label{pimdef2}
\end{align}
We need to apply the variable change
\begin{align}
\sigma_x:(x,y)\rightarrow(x^{-1},xy)
\end{align}
to translate the index of the boundary theory to the
index on the theory on the $R_x$-fixed locus \cite{Arai:2019xmp,Arai:2020qaj,Gaiotto:2021xce}.
Namely, ${\cal F}_{m,0}$ should be given by ${\cal F}_{m,0}=\sigma_x\Pi_m$.
With the analytic expression (\ref{pimdef2})
for $\Pi_m$ it is easy to calculate $\sigma_x\Pi_m$ explicitly.
For small $m$ the explicit forms of $t$-expansion of ${\cal F}_{m,0}$ are
\begin{align}
{\cal F}_{1,0}
&=\frac{-x}{1-\frac{y}{x}}
+\frac{x^2(-1+\frac{y}{x}-\frac{y^2}{x^2})}{1-\frac{y}{x}}
+\frac{x^3(-1+\frac{y}{x}-\frac{y^2}{x^2})}{1-\frac{y}{x}}
+\frac{x^4(-1+\frac{y}{x}-\frac{y^2}{x^2})}{1-\frac{y}{x}}
+{\cal O}(t^5),\nonumber\\
{\cal F}_{2,0}
&=\frac{-\frac{x^5}{y}}{1-\frac{y^2}{x^2}}
+\frac{\frac{x^7}{y^2}(-1+\frac{y^2}{x^2}-\frac{y^3}{x^3})}{1-\frac{y}{x}}
+\frac{\frac{x^9}{y^3}(-1-\frac{y}{x}-\frac{y^2}{x^2}+\frac{y^3}{x^3}-\frac{y^5}{x^5}-\frac{y^8}{x^8})}{1-\frac{y^2}{x^2}}
+{\cal O}(t^7),\nonumber\\
{\cal F}_{3,0}
&=\frac{-\frac{x^{12}}{y^3}}{1-\frac{y^3}{x^3}}
+\frac{\frac{x^{15}}{y^5}(-1-\frac{y^2}{x^2}+\frac{y^3}{x^3}-\frac{y^6}{x^6})}{1-\frac{y^3}{x^3}}
+\frac{\frac{x^{18}}{y^7}(-1-\frac{y^2}{x^2}-2\frac{y^4}{x^4}+\frac{y^5}{x^5}-\frac{y^8}{x^8}-\frac{y^{12}}{x^{12}})}{1-\frac{y^3}{x^3}}
+{\cal O}(t^{12})
\end{align}
${\cal F}_{0,m}$ are obtained from ${\cal F}_{m,0}$ by exchanging $x$ and $y$;
${\cal F}_{0,m}={\cal F}_{m,0}|_{x\leftrightarrow y}$.

If both $m_x$ and $m_y$ are positive, we naively expect that
${\cal F}_{m_x,m_y}$ is the index of
a $U(m_x)\times U(m_y)$ gauge theory with
bi-fundamenal fields living on the intersection ${\cal D}_x\cap{\cal D}_y$.
The intersection is a segment, and
we need to carefully determine
the boudnary conditions at the ends of the segment
to calculate the letter index of the bi-fundamental fields.
We will not examine the boundary conditions here.
In fact, numerical analysis shows that
only ${\cal F}_{m_x,0}$ and ${\cal F}_{0,m_y}$ are sufficient
to reprodcue the finite $k$ index in (\ref{kpart}),
and we do not have to include the contributions from
intersecting branes.
This means ${\cal F}_{m_x,m_y}=0$ for $m_x,m_y\geq1$,
and strongly suggests that the supersymmetry is spontaneously broken
in the theory realized on ${\cal D}_x\cup{\cal D}_y$.

In summary, the brane expansion of $I_{\infty,k}$ is
\begin{align}
\frac{I_{\infty,k}}{I_{\rm sugra}I_{\rm D5}}
=1+\sum_{m=1}^\infty x^{mk}{\cal F}_{m,0}
+\sum_{m=1}^\infty y^{mk}{\cal F}_{0,m}.
\label{fink2}
\end{align}
We numerically confirmed that this relation holds for $k\leq20$ up to ${\cal O}(t^{20})$.

\subsection{Reduced expansion}
For the Schur index without line insertion
the double sum expansion in \cite{Arai:2020qaj} is reduced to
simple-sum expansion
if we treat the index as a $y$-series \cite{Gaiotto:2021xce}.
See also \cite{Imamura:2022aua,Fujiwara:2023bdc} for the mechanism of the reduction.
This reduction mechanism works for the brane expansion of the
line-operator index (\ref{fink2}).
We can easily confirm that $y$-series expansion of ${\cal F}_{0,m}=(\sigma_x\Pi_m)|_{x\leftrightarrow y}$ ($m\geq1$)
is zero, and (\ref{fink2}) reduces to the simple-sum expansion
\begin{align}
\frac{I_{\infty,k}}{I_{\rm sugra}I_{\rm D5}}=\sum_{m=0}^\infty x^{km}{\cal F}_{m,0}.
\label{finiteksimple}
\end{align}
${\cal F}_{m,0}$ expressed as $y$-series are shown for small $m$ below.
\begin{align}
{\cal F}_{1,0}&=\frac{-x}{1-x}-y+(-\tfrac{1}{x}-1)y^2+(-\tfrac{1}{x^2}-\tfrac{1}{x})y^3+{\cal O}(y^4),\nonumber\\
{\cal F}_{2,0}&=\frac{x^3}{(1-x)(1-x^2)}+\frac{x}{1-x}y+\frac{\frac{1}{x}+x}{1-x}y^2+\frac{\frac{1}{x^3}+\frac{1}{x}}{1-x}y^3+{\cal O}(y^4),\nonumber\\
{\cal F}_{3,0}
&=\frac{-x^6}{(1-x)(1-x^2)(1-x^3)}
+\frac{-x^3}{(1-x)(1-x^2)}y
+\frac{\-1+x-x^2}{(1-x)^2}y^2
+{\cal O}(y^3).
\label{fm0}
\end{align}
We numerically confirmed that
(\ref{finiteksimple}) holds
for $k\leq 20$ up to ${\cal O}(y^{10})$ and ${\cal O}(x^{20})$.

\subsection{Half BPS limit}
We can simplify the problem by taking
the half-BPS limit $y\rightarrow 0$.
Then, functions appearing in (\ref{finiteksimple}) take simple forms
\begin{align}
\frac{I_{\infty,n}}{I_{\rm sugra}}=\Pi_n=\Pexp(x+\cdots+x^n),\quad
I_{\rm D5}=I_{\rm sugra}=\Pexp\frac{x}{1-x},
\end{align}
and
we can analytically confirm (\ref{finiteksimple}).
In fact, the simple-sum brane expansion (\ref{finiteksimple})
is identical with the simple-sum giant graviton expansion of
the Schur index without line insertion \cite{Gaiotto:2021xce}
in the half BPS limit.

\section{Finite $N$ corrections}
\subsection{Full expansion}
If $N$ is finite, in addition to disk D3-branes on ${\cal D}_x$ and ${\cal D}_y$,
ones on ${\cal D}'_x$ and ${\cal D}'_y$ also contribute to the index.
We expect the quadruple-sum brane expansion
\begin{align}
\frac{I_{N,k}}{I_{\rm sugra}I_{\rm D5}}
=\sum_{m_x,m_y,m'_x,m'_y=0}^\infty
x^{km_x+k'm'_x}y^{km_y+k'm'_y}{\cal F}_{m_x,m_y,m'_x,m'_y},
\label{finiten}
\end{align}
where $k'=N-k$.
Functions
${\cal F}_{m_x,m_y,m'_x,m'_y}$
are not all independent but satisfy
\begin{align}
{\cal F}_{m_x,m_y,m'_x,m'_y}
={\cal F}_{m'_x,m'_y,m_x,m_y}
={\cal F}_{m_y,m_x,m'_y,m'_x}|_{x\leftrightarrow y}
\label{frelations}
\end{align}
and by definition
\begin{align}
{\cal F}_{m_x,m_y,0,0}={\cal F}_{m_x,m_y}.
\end{align}
We have not yet fully understood how to calculate the
functions ${\cal F}_{m_x,m_y.m'_x,m'_y}$
for general wrapping numbers.
In this subsection, we will calculate the $t$-series expansion of
$I_{N,k}$ on the gauge theory side for different values of $N$ and $k$, and
we extract functions ${\cal F}_{m_x,m_y,m'_x,m'_y}$ for small wrapping numbers
by assuming the expansion (\ref{finiten}).

Let $m_{\rm tot}$ be the total wrapping number; $m_{\rm tot}=m_x+m_y+m'_x+m'_y$.
There are four functions for $m_{\rm tot}=1$,
and they are all related
to
${\cal F}_{1,0,0,0}={\cal F}_{1,0}=\sigma_x\Pi_1$
by (\ref{frelations}).

For $m_{\rm tot}=2$,
there are four essentially different functions
\begin{align}
{\cal F}_{2,0,0,0},\quad
{\cal F}_{1,1,0,0},\quad
{\cal F}_{1,0,1,0},\quad
{\cal F}_{1,0,0,1}.
\end{align}
The first two are the same as the functions given in the previous section
\begin{align}
{\cal F}_{2,0,0,0}&={\cal F}_{2,0}=\sigma_x\Pi_2,\nonumber\\
{\cal F}_{1,1,0,0}&={\cal F}_{1,1}=0,
\end{align}
and the other two extracted from the comparison to $I_{N,k}$ are
\begin{align}
{\cal F}_{1,0,1,0}
&=\frac{x^2(1+\frac{y}{x})}{(1-\frac{y}{x})^2}
+\frac{x^3(2-2\frac{y}{x}+2\frac{y^2}{x^2}+2\frac{y^3}{x^3})}{(1-\frac{y}{x})^2}
\nonumber\\&\quad
+\frac{x^4(3-4\frac{y}{x}+4\frac{y^2}{x^2}-6\frac{y^3}{x^3}+6\frac{y^4}{x^4}+3\frac{y^5}{x^5})}{(1-\frac{y}{x})^2}
+{\cal O}(t^5),\nonumber \\
{\cal F}_{1,0,0,1}
&=\frac{2xy}{(1-\frac{y}{x})(1-\frac{x}{y})}
+\frac{2(x^3+y^3)}{(1-\frac{y}{x})(1-\frac{x}{y})}
\nonumber\\&\quad
+\frac{2(\frac{x^5}{y}+x^4-x^3y+x^2y^2-xy^3+y^4+\frac{y^5}{x})}{(1-\frac{y}{x})(1-\frac{x}{y})}
+{\cal O}(t^5).
\label{f11}
\end{align}
These are expected to be the indices of
$U(1)\times U(1)$ gauge theories with bi-fundamental fields.

The brane system corresponding to ${\cal F}_{1,0,1,0}$ consists of two disks 
${\cal D}_x$ and ${\cal D}'_x$, and
they are adjacent on ${\cal D}_x\cap{\cal D}'_x=S^2$.
There exists a bi-fundamenal hypermultiplet on $S^2$.
Indeed, we can confirm that ${\cal F}_{1,0,1,0}$
in (\ref{f11}) is given by
\begin{align}
{\cal F}_{1,0,1,0}={\cal F}_{1,0}^2\sigma_xI_{\rm 3d},
\label{f1010}
\end{align}
where
$I_{\rm 3d}$ is the 3d index \cite{Kim:2009wb,Imamura:2011su} of a charged hypermultiplet
\begin{align}
I_{\rm 3d}=\int\frac{dz}{2\pi iz}
\Pexp\left[
i_{\rm 3d}\left(z+\frac{1}{z}\right)
\right],\quad
i_{\rm 3d}=\frac{y^{\frac{1}{2}}(1-x)}{1-q}.
\end{align}

${\cal F}_{1,0,0,1}$ is the index of the theory realized on
${\cal D}_x\cup{\cal D}'_y$.
Two disks ${\cal D}_x$ and ${\cal D}'_y$ are touching each other
at two points, and bi-fundamental
fields live on the points.
We find the relation
\begin{align}
{\cal F}_{1,0,0,1}=2{\cal F}_{1,0}{\cal F}_{0,1}
\end{align}
holds, and identify the factor $2$ as the contribution
from the bi-fundamental fields.
Indeed, a pair of bi-fundamental fermion with
the letter index $-(z+\frac{1}{z})$ gives this factor
\begin{align}
2=\int\frac{dz}{2\pi iz}
\Pexp\left[
-\left(z+\frac{1}{z}\right)
\right].
\end{align}

Using these functions ${\cal F}_{m_x,m_y,m'_x,m'_y}$ for $m_{\rm tot}\leq 2$,
the quadruple-sum expansion (\ref{finiten})
correctly reproduces $I_{N,k}$ up to order ${\cal O}(t^n)$ with
$n=\min(2k+k',k+2k')$.

For $m_{\rm tot}=3$,
there are five essentially different functions:
\begin{align}
{\cal F}_{3,0,0,0},\quad
{\cal F}_{2,1,0,0},\quad
{\cal F}_{2,0,1,0},\quad
{\cal F}_{2,0,0,1},\quad
{\cal F}_{1,1,1,0}.
\end{align}
The first two are given by ${\cal F}_{3,0,0,0}={\cal F}_{3,0}=\sigma_x\Pi_3$ and
${\cal F}_{2,1,0,0}={\cal F}_{2,1}=0$,
and we can extract the other three by assuming the quadruple-sum expansion (\ref{finiten}).
The leading terms in the $t$-expansions
are
\begin{align}
{\cal F}_{2,0,1,0}&=\frac{x^4}{(1-\frac{y}{x})^2(1-\frac{y^2}{x^2})}+{\cal O}(t^5),\nonumber\\
{\cal F}_{2,0,0,1}&=\frac{-x^3y}{(1-\frac{y}{x})^2(1-\frac{y^2}{x^2})}+{\cal O}(t^5),\nonumber\\
{\cal F}_{1,1,1,0}&=\frac{-xy^3}{(1-\frac{y}{x})^3}+{\cal O}(t^5).
\label{f3}
\end{align}
Although it would also be possible to obtain higher order terms
in (\ref{f3}) and functions for $m_{\rm tot}\geq4$ by continuing a similar analysis,
we stop here the analysis of the quadruple-sum expansion (\ref{finiten}).

\subsection{Reduced expansion}
We can reduce the quadruple-sum expansion (\ref{finiten}) to
the double-sum expansion
\begin{align}
\frac{I_{N,k}}{I_{\rm sugra}I_{\rm D5}}
=\sum_{m,m'=0}^\infty
x^{km+k'm'}{\cal F}_{m,0,m',0},
\label{finitendbl}
\end{align}
by treating the index as a $y$-series.
In this expansion only the $R_x$-fixed locus ${\cal C}_x$
contributes to the index,
and the summation is taken over the two wrapping numbers
$m$ and $m'$ asscoiated with ${\cal D}_x$ and ${\cal D}'_x$,
respectively.

${\cal F}_{m,0,m',0}$ is the index of the theory on ${\cal C}_x$
divided by the D5-brane to ${\cal D}_x$ and ${\cal D}'_x$.
We take the S-duality frame in which the tubuler D5-brane
becomes an NS5-brane.
Then, the gauge theory on ${\cal C}_x$
is the quiver gauge theory with gauge group $U(m)\times U(m')$.
The vector multiplets live in two disks ${\cal D}_x$ and ${\cal D}'_x$,
and
the bi-fundamental hypermultiplet
lives on the wall dividing ${\cal C}_x$ into the two parts.
The functions ${\cal F}_{m,0,m',0}$ are given by the localization formula
\begin{align}
{\cal F}_{m,0,m',0}
&=
\sigma_x
\Bigg[
\int_{U(m)}\hspace{-1.3em}dU
\int_{U(m')}\hspace{-1.3em}dU'
\Pexp\bigg(
i_{\rm half}\chi_{\rm adj}^{U(m)}(U)
\nonumber\\&\quad\quad
+i_{\rm half}\chi_{\rm adj}^{U(m')}(U')
+i_{\rm 3d}\chi^{U(m)\times U(m')}_{\rm bifund}(U,U')
\bigg)
\Bigg],
\end{align}
where the bi-fundamental character is defined by
\begin{align}
\chi^{U(m)\times U(m')}_{\rm bifund}(U,U')
=\tr(U)\tr(U'^{-1})
+\tr(U^{-1})\tr(U').
\end{align}
The first few terms in the $y$-series expansions of the functions for small $m_{\rm tot}$ are
\begin{align}
{\cal F}_{1,0,1,0}
&=\frac{x^2}{{(1-x)}^{2}}
+\frac{x(3-x)}{1-x}y
+\frac{5-x+ x^2 -x^3}{1-x}y^2
+\mathcal{O}(y^3),
\nonumber\\
{\cal F}_{2,0,1,0}
&=\frac{-x^4}{(1-x)^2(1-x^2)}
+\frac{x^2(-2-2x+x^2)}{(1-x)(1-x^2)}y
\nonumber\\&\quad
+\frac{-2-2x+2x^2-2x^3+x^4}{(1-x)^2}y^2
+\mathcal{O}(y^3),
\nonumber\\
{\cal F}_{3,0,1,0}
&=\frac{x^7}{(1-x)^2(1-x^2)(1-x^3)}
+\frac{x^4(2+x+2x^2-x^3)}{(1-x)(1-x^2)(1-x^3)}y
\nonumber\\&\quad
+\frac{x(2+2x+3x^2-x^3+2x^4+x^6-x^7)}{(1-x)(1-x^2)(1-x^3)}y^2
+{\cal O}(y^3),
\nonumber\\
{\cal F}_{2,0,2,0}
&=\frac{x^6}{(1-x)^2(1-x^2)^2}
+\frac{x^3(1+2x-x^2)}{(1-x)^2(1-x^2)}y
\nonumber\\&\quad
+\frac{2x(1+3x-3x^2+2x^4-x^5)}{(1-x)^2(1-x^2)}y^2
+{\cal O}(y^3).
\end{align}
See (\ref{fm0}) for ${\cal F}_{m,0,0,0}={\cal F}_{m,0}=\sigma_x\Pi_m$.

We numerically confirmed
the expansion (\ref{finitendbl}) works well.
For example, for $N=2$ and $k=1$, if we sum up the contributions from $m_{\rm tot}\leq4$,
the discrepancy is
\begin{align}
\frac{I_{2,1}}{I_{\rm sugra}I_{\rm D5}}-\sum_{m+m'\leq4}(\cdots)
={\cal O}({x}^{14})y^0
+{\cal O}({x}^{10})y^1
+{\cal O}(x^7)y^2
+{\cal O}(x^4)y^3
+{\cal O}(y^4),
\end{align}
and a lot of terms are correctly reproduced.

\subsection{Half BPS limit}
The half-BPS limit $y\rightarrow0$ of $I_{N,k}$ for general values of
$N$ and $k$ is given by \cite{Hatsuda:2023imp}
\begin{align}
I_{N,k}=
\prod_{m=1}^k\frac{1}{1-x^m}
\prod_{m=1}^{N-k}\frac{1}{1-x^m}.
\label{halfink}
\end{align}
The expansion (\ref{finitendbl}) reproduces this factorized form.
In the half-BPS limit the letter index $\sigma_xi_{\rm 3d}$ of the hypermultiplet on the wall vanishes,
and the wall contribution becomes trivial.
As the result, the functions ${\cal F}_{m,0,m',0}$ are
factorized to the product of ${\cal F}_{m,0}$ and ${\cal F}_{m',0}$;
\begin{align}
{\cal F}_{m,0,m',0}={\cal F}_{m,0}{\cal F}_{m',0}=
\sigma_x(\Pi_m\Pi_{m'}).
\end{align}
Then, the double-sum expansion
(\ref{finitendbl}) factorizes into two copies of the
simple-sum expansion (\ref{finiteksimple}) for the large $N$ index
\begin{align}
\frac{I_{N,k}}{I_{\rm sugra}I_{\rm D5}}
&=\left(\sum_{m=0}^\infty
x^{km}\sigma_x\Pi_m\right)
\left(\sum_{m'=0}^\infty
x^{k'm'}\sigma_x\Pi_{m'}\right)
=\frac{I_{\infty,k}}{I_{\rm sugra}I_{\rm D5}}\frac{I_{\infty,k'}}{I_{\rm sugra}I_{\rm D5}}.
\label{finitendb2}
\end{align}
This correctly reproduces the factorized form
(\ref{halfink}).

\section{Conclusions and discussion}\label{disc.sec}
In this paper we investigated the brane expansions for the index
$I_{N,k}$ of the ${\cal N}=4$ $U(N)$ SYM with
the line operators of the rank $k$ anti-symmetric representations.
Our analysis was based on the D5-brane realization of the line-operators.

The large $N$ index $I_{\infty,k}$ with finite $k$
is given by the double-sum brane expansion (\ref{finitek}),
and the finite $N$ index $I_{N,k}$ is given by the
quadruple-sum brane expansion (\ref{finiten}).
By taking an appropriate expansion scheme they reduce to
the simple-sum expansion (\ref{finiteksimple}) and
the double-sum expansion (\ref{finitendbl}), respectively.
The objects causing
the finite $k$ and finite $N$ corrections
are disk D3-branes ending on the D5-brane.

For the quadruple-sum expansion (\ref{finiten}) for finite $N$,
the brane indices ${\cal F}_{m_x,m_y,m'_x,m'_y}$ are labeled by
the four wrapping numbers,
and the corresponding brane systems consist of
branes extended on the four disks ${\cal D}_x$, ${\cal D}_y$, ${\cal D}'_x$, and ${\cal D}'_y$.
Due to the complication of the systems,
we did not give a general formula
to calculate such contributions.
One interesting fact we found
is the contributions ${\cal F}_{m_x,m_y}={\cal F}_{m_x,m_y,0,0}$
for $m_x,m_y\geq1$
are vanishing.
This indicates the supersymmetry is
spontaneously broken in the theory realized on
${\cal D}_x\cup{\cal D}_y$.
In connection with this,
another interesting fact we found is
${\cal F}_{1,1,1,0}\neq0$.
This means that
the corresponding brane system
${\cal D}_x\cup{\cal D}_y\cup{\cal D}'_x$
preserves supersymmetry even though
it includes the supersymmetry breaking configuration
${\cal D}_x\cup{\cal D}_y$ as a subsystem.
This indicates that whether the supersymmetry is broken
depends on the global structure of the brane system.
This situation reminds us of the $s$-rules
in Hanany-Witten type brane systems \cite{Hanany:1996ie}.
It would be interesting to investigate the
supersymmetry breaking/preserving mechanism on the
brane systems with general wrapping numbers.

There are different directions of generalization
of the results in this work.

One is generalization to multiple tubular D5-branes.
In this work we focused on the anti-symmetric Wilson lines
with $\mu=\{1^k\}$.
We can consider a more general
line labeled by $\mu=\{n^{k_1-k_2},(n-1)^{k_2-k_3},\ldots,1^{k_n}\}$.
The Young diagram has $n$ columns, and the line corresponds to
concentric $n$ D5-branes.
The $n$ D5-branes divide each of ${\cal C}_x$ and ${\cal C}_y$
into $n+1$ zones.
If we treat the index as $y$-series,
only branes on ${\cal C}_x$ will contribute to the
index, and we will have a gauge theory depicted as
a linear quiver with $n+1$ nodes.
The full expansion will be more complicated and
the theory on a brane system is in general
a quiver gauge theory
with $2(n+1)$ nodes.

It is also important to find precise line operators
corresponding to tubular D3-branes, and investigate the finite $k$ and finite $N$
corrections to the corresponding line operator index.
Although it is often claimed in the literature that
the line operator in a symmetric representation
is holographically dual to a tubular D3-brane configuration found in [6],
fluctuation modes on the D3-brane analyzed in \cite{Faraggi:2011bb}
do not reproduce
the index of the symmetric line operators,
which is identical in the large $N$ limit to the index (11)
of the anti-symmetric line operator with the same rank \cite{Hatsuda:2023iwi}.
(See Appendix \ref{d3.sec}.)
This means some modification is necessary
for line operators dual to the tubular D3-branes.
Regardless of whatever correspond to tubular D3-branes,
it would be interesting to analyze
the line-operator index for the tubular D3-brane insertion
by using brane expansions.
We point out that D-strings are strong candidates for
causing finite $k$ corrections.
Let us introduce a coordinate $u$ in $AdS_5$
so that the metric of $AdS_5$ is given by
\begin{align}
ds_{AdS_5}^2=L_{\rm AdS}^2(\cosh^2u ds_{AdS_2}^2
+du^2
+\sinh^2u ds_{S^2}^2).
\end{align}
For a tubular D3-brane with string charge $k$, its worldvolume
is given by $u=u_{\rm D3}$ \cite{Drukker:2005kx} where
$u_{\rm D3}$ is given by
\begin{align}
\sinh u_{\rm D3}=\frac{k\sqrt{4\pi g_sN}}{4N}.
\end{align}
The mass of a D-string stretched between the two poles of $S^2$ is
\begin{align}
T_{\rm D1}\int_{-u_{\rm D3}}^{u_{\rm D3}} L_{\rm AdS}\cosh u du
=2T_{\rm D1}L_{\rm AdS}\sinh u_{\rm D3}
=\frac{k}{L_{\rm AdS}},
\end{align}
and the dimension of the corresponding operator is $k$.
This fact strongly suggests that D-strings
cause finite $k$ corrections to the line operator index.
In addition, D3 giants will give finite $N$ corrections.
It would be interesting
to study brane expansions including the contribution from
D-strings and D3 giants.

It is expected that when the numbers of columns and rows
of Young diagrams become of order $N$
the holographic correspondent should be
bubbling geometry \cite{DHoker:2007mci}.
Recently, relation between the Schur indices of line operators
with such large representations and bubbling geometries
were discussed in \cite{Hatsuda:2023iof}.
It is interesting to investigate the finite $N$ corrections
in such background geometries
by using brane expansions.

We hope to revisit these issues in future works.

\section*{Acknowledgments}
The authors thank A.~Sei and D.~Yokoyama for valuable discussions and comments.
The work of Y.~I. was supported by JSPS KAKENHI Grant Number JP21K03569.
The work of M.~I. was supported by JST SPRING, Grant Number JPMJSP2106.

\appendix
\section{Character expansion method}\label{characterexp.sec}
In this appendix we derive some useful formulas
using character expansion method
\cite{Dolan:2007rq,Dutta:2007ws}.
Let us introduce the power sum symmetric polynomials
\begin{align}
p_\lambda(U)=\prod_{i=1}^{\ell(\lambda)}\tr U^{\lambda_i}
=[\tr(U)]_\lambda
\end{align}
labeled by partitions $\lambda$,
and define the index with insertion of line operators associated with these polynomials.
We start from the localization formula
(\ref{localization}) with $s_\mu$ and $s_\nu$ replaced by $p_\lambda$ and $p_{\lambda'}$:
\begin{align}
I_{N,\lambda,\lambda'}
&=\langle p_\lambda(U)p_{\lambda'}(U^{-1})\rangle^{U(N)}
\nonumber\\
&=\int_{U(N)}\hspace{-1.3em}dU
\Pexp(i_{\rm vec}\tr(U)\tr(U^{-1}))p_\lambda(U)p_{\lambda'}(U^{-1}).
\end{align}
By using the definition of the plethystic exponential we can rewrite this
as
\begin{align}
I_{N,\lambda,\lambda'}
=
\int_{U(N)}\hspace{-1.3em}dU
\sum_{\lambda''}\frac{1}{z_{\lambda''}}
[i_{\rm vec}]_{\lambda''}
p_{\lambda+\lambda''}(U)p_{\lambda'+\lambda''}(U^{-1}),
\end{align}
where $\lambda+\lambda'$ is the union of two partitions $\lambda$ and $\lambda'$,
which is given by arranging $\{\lambda_1,\lambda_2,\ldots,\lambda_{\ell(\lambda)},\lambda'_1,\lambda'_2,\ldots,\lambda'_{\ell(\lambda')}\}$
in descending order.
The polynomials $p_\lambda(U)$ are expressed as linear combinations of the Schur polynomials $s_\mu(U)$
by the formula
\begin{align}
p_\lambda(U)=\sum_{\mu,\ell(\mu)\leq N}\chi_\mu(\lambda)s_\mu(U)
\label{frobenius}
\end{align}
given by Frobenius.
$\chi_\mu(\lambda)$ is the character of
an irreducible representation of the symmetric group $S_n$ ($n=|\lambda|$) corresponding to the
partition $\mu$ evaluated at a permutation of cycle type $\lambda$.
With this formula and the orthonormal relation of
the Schur polynomials
\begin{align}
\int_{U(N)}\hspace{-1.3em}dU s_\mu(U)s_\nu(U^{-1})=\delta_{\mu,\nu},
\end{align}
we obtain the expression
\begin{align}
I_{N,\lambda,\lambda'}
=
\sum_{\lambda''}\frac{1}{z_{\lambda''}}
[i_{\rm vec}]_{\lambda''}
\sum_{\mu,\ell(\mu)\leq N}
\chi_\mu(\lambda+\lambda'')\chi_\mu(\lambda'+\lambda'').
\label{illformula}
\end{align}

In the large $N$ limit,
the bound $\ell(\mu)\leq N$ in the $\mu$-sum
in (\ref{illformula}) dissapears, and
we can use the orthogonal relation of the characters $\chi_\mu(\lambda)$:
\begin{align}
\sum_\mu\chi_\mu(\lambda)\chi_\mu(\lambda')=z_\lambda\delta_{\lambda,\lambda'}.
\end{align}
Then, we obtain
\begin{align}
I_{\infty,\lambda,\lambda'}
&=
\delta_{\lambda,\lambda'}\sum_{\lambda''}\frac{z_{\lambda+\lambda''}}{z_{\lambda''}}
[i_{\rm vec}]_{\lambda''}
=z_\lambda\delta_{\lambda,\lambda'}[I_{\rm F1}]_\lambda I_{\rm sugra}.
\label{ill2}
\end{align}
For $\lambda=\lambda'=\{1\}$ this gives (\ref{Ifund}).

By using the inverse relation to 
(\ref{frobenius})
\begin{align}
s_\mu(U)=\sum_{\lambda\vdash|\mu|} \frac{1}{z_\lambda}\chi_\mu(\lambda)p_\lambda(U),
\label{stot}
\end{align}
We can express $\langle s_\mu(U)s_\nu(U^{-1})\rangle^{U(N)}$ as a
linear combination of $I_{N,\lambda,\lambda'}$.
For $N=\infty$,
combining 
(\ref{ill2}) and
(\ref{stot})
we obtain
\begin{align}
\langle s_\mu(U)s_\nu(U^{-1})\rangle^{U(\infty)}
&=I_{\rm sugra}\sum_{\lambda\vdash|\mu|}\frac{1}{z_\lambda}
[I_{\rm F1}]_\lambda\chi_\mu(\lambda)\chi_\nu(\lambda).
\label{largenmn}
\end{align}
This is non-vanishing only when $|\mu|=|\nu|$.

The line operator indices for the symmetric and anti-symmetric representations
can be obtained by substituting
$\sigma_{\{1^k\}}(\lambda)=\pm 1$
and $\sigma_{\{k\}}(\lambda)=1$ for $\lambda\vdash k$
into (\ref{largenmn}).
The common result is shown in (\ref{kpart}).

\section{Letter index from a D3-brane on $AdS_2\times S^2$}\label{d3.sec}
Fluctuation modes on a D3-branes extending along $AdS_2\times S^2$
were studied in \cite{Faraggi:2011bb} in detail.
They belong to the $OSp(4^*|4)$
representation
\begin{align}
\bigoplus_{j=0,1,2,\ldots}\bm{j},
\end{align}
where $\bm{j}$ are $OSp(4^*|4)$ short multiplets whose decomposition
to $so(2,1)\times so(3)\times so(5)$ irreducible representations are
\begin{align}
\bm{j}
&=(j+1,j,\bm{5})\oplus
(j+\tfrac{3}{2},j+\tfrac{1}{2},\bm{4})\oplus
(j+2,j+1,\bm{1})
\nonumber\\&\quad
\oplus
(j+\tfrac{1}{2},j-\tfrac{1}{2},\bm{4})\oplus
(j+1,j,\bm{1})
\end{align}
for $\bm{j}\geq1$ and
\begin{align}
\bm{0}
=(1,0,\bm{5})
\oplus(\tfrac{3}{2},\tfrac{1}{2},\bm{4})
\oplus(2,1,\bm{1}).
\end{align}
$(d,\ell,\bm{r})$
is the tensor product of the $so(2,1)$ (conformal) representation
with the primary state with $H=d$,
the $so(3)$ spin $\ell$ representation with
the weights $J_1=\ell,\ell-1,\ldots,-\ell$,
and one of the $so(5)$ representations
\begin{align}
\bm{r}=\bm{1}:(R_x,R_y)&=(0,0),\nonumber\\
\bm{r}=\bm{4}:(R_x,R_y)&=(\pm\tfrac{1}{2},\pm\tfrac{1}{2}),\nonumber\\
\bm{r}=\bm{5}:(R_x,R_y)
&=(\pm1,0),(0,\pm1),(0,0).
\end{align}
The quantum numbers of the BPS states in $\bm{j}$ saturating the bound $H\geq J_1+R_x+R_y$ are
\begin{align}
(J_1,R_x,R_y)=(j,1,0),(j,0,1),(j+\tfrac{1}{2},\tfrac{1}{2},\tfrac{1}{2}),(j-\tfrac{1}{2},\tfrac{1}{2},\tfrac{1}{2}),
\end{align}
for $j\geq1$ and
\begin{align}
(J_1,R_x,R_y)=(0,1,0),(0,0,1),(\tfrac{1}{2},\tfrac{1}{2},\tfrac{1}{2}).
\end{align}
for $j=0$.
Based on the definition of the index (\ref{indexdef}),
we obtain the following contributions to the letter index $i_{\rm D3}$
from the irreducible $OSp(4^*|4)$ representations $\bm{j}$:
\begin{align}
j\geq1 &: q^jx+q^jy
-q^{j+\frac{1}{2}}x^{\frac{1}{2}}y^{\frac{1}{2}}
-q^{j-\frac{1}{2}}x^{\frac{1}{2}}y^{\frac{1}{2}}
=-q^j(1-x)(1-y),\nonumber\\
j=0 &: x+y-q^{\frac{1}{2}}x^{\frac{1}{2}}y^{\frac{1}{2}}
=1-(1-x)(1-y),
\end{align}
where $q=xy$.
By summing up all the contributions we obtain
\begin{align}
i_{\rm D3}=1-\frac{(1-x)(1-y)}{1-xy}.
\end{align}
(Interestingly, this coincides with the letter index of the
${\cal N}=4$ vector multiplet (\ref{ivec}).)
$\Pexp i_{\rm D3}$
does not agree with the result for the symmetric representation
on the gauge theory side \cite{Hatsuda:2023iwi}
\begin{align}
\lim_{k\rightarrow\infty}\frac{1}{I_{\rm sugra}}
\langle s_{\{k\}}(U)s_{\{k\}}(U^{-1})\rangle^{U(\infty)}=\Pexp\left(\frac{x}{1-x}+\frac{y}{1-y}\right).
\end{align}

\end{document}